\documentstyle[preprint,pra,eqsecnum,aps]{revtex}
\begin{document}
\draft

\title{$O(3,3)$-like Symmetries of Coupled Harmonic Oscillators}

\author{D. Han\footnote{electronic mail: han@trmm.gsfc.nasa.gov}}
\address{National Aeronautics and Space Administration, Goddard Space
Flight Center, Code 910.1, Greenbelt, Maryland 20771}

\author{Y. S. Kim\footnote{electronic mail: kim@umdhep.umd.edu}}
\address{Department of Physics, University of Maryland, College Park,
Maryland 20742}

\author{Marilyn E. Noz \footnote{electronic mail: noz@nucmed.med.nyu.edu}}
\address{Department of Radiology, New York University, New York, New York
10016}

\maketitle


\begin{abstract}
In classical mechanics, the system of two coupled harmonic oscillators
is shown to possess the symmetry of the Lorentz group $O(3,3)$ applicable
to a six-dimensional space consisting of three space-like and three
time-like coordinates, or $SL(4,r)$ in the four-dimensional phase space
consisting of two position and two momentum variables.  In quantum
mechanics, the symmetry is reduced to that of $O(3,2)$ or $Sp(4)$, which
is a subgroup of $O(3,3)$ or $SL(4,r)$ respectively.  It is shown that
among the six $Sp(4)$-like subgroups, only one possesses the symmetry
which can be translated into the group of unitary transformations in
quantum mechanics.  In quantum mechanics, there is the lower bound in the
size of phase space for each mode determined by the uncertainty principle
while there are no restriction on the phase-space size in classical
mechanics.  This is the reason why the symmetry is smaller in quantum
mechanics.

\end{abstract}


\section{Introduction}\label{intro}
For two coupled harmonic oscillators, there is a tendency to believe that
the problem is completely and thoroughly understood at the level of
Goldstein's textbook on classical mechanics \cite{gold80} and
that no further studies are necessary.	We start this paper with the
following prejudices.

\begin{itemize}

\item[(a)] The group $O(3,3)$ or $SL(4,r)$ is only of mathematical interest,
and does not appear to possess any relevance to the physical world, although
the fifteen Dirac matrices in the Majorana representation constitute the
generators of $O(3,3)$ \cite{dir63,lee95}.

\item[(b)] The transition from classical to quantum mechanics of this
oscillator system is trivial once the problem is brought to a diagonal form
with the appropriate normal modes.

\item[(c)] The diagonalization requires only a rotation, and no other
transformations are necessary.

\end{itemize}

It is known but not widely known that the diagonalization requires squeeze
transformations in addition to the rotation \cite{arn78,abra78,guil84}.
This means that the diagonalization process in classical mechanics is a
symplectic transformation.  It is also known that symplectic
transformations perform linear canonical transformations in classical
mechanics which can be translated into unitary transformations in quantum
mechanics~\cite{arn78,abra78,knp91}.

In this paper, we are interested in the size of phase space.  In classical
mechanics, the size can grow or shrink.  We shall show first in this paper
that, in classical mechanics, the symmetry group for the two-coupled
harmonic oscillators becomes $SL(4,r)$ which is locally isomorphic to $O(3,3)$.
This group is not symplectic but has a number of symplectic subgroups which
are locally isomorphic to $O(3,2)$ or $O(2,3)$.  This conclusion is quite
different from our earlier contention that the $O(3,2)$ will solve all the
problems for the two-mode oscillator system \cite{hkny93}.

Let us translate the above mathematics into the language of physics.  In
quantum mechanics, the size of phase space is allowed to grow resulting in
an increased entropy, but it cannot shrink beyond the limit imposed by the
uncertainty principle \cite{knp91}.  If we apply this principle to each of
the normal modes, only the $O(3,2)$ symmetry is carried into quantum
mechanics.

It is known also that not all the transformations in quantum mechanics are
unitary, and the expansion of phase space has its well-defined place in
the quantum world \cite{fey72,ume82,yupo87,ekn89}.  The problem is the
shrinking of phase space.  In the system of two coupled oscillators, the
phase space expansion in one mode means the shrinking phase space in the
other mode.  While we do not provide a complete solution to the problem,
we can give a precise statement of the issue in this paper.

The coupled oscillator problem is covered in freshman physics.  It
stays with us in many different forms because it provides the mathematical
basis for many soluble models in physics, including the Lee model in
quantum field theory \cite{tdlee54,sss61}, the Bogoliubov transformation
in superconductivity \cite{bogo58,fewa71,tink75}, relativistic models of
elementary particles \cite{knp91,van85}, and squeezed states of light
\cite{dir63,yuen76,cav85,bishop88}.  In this paper, in addition to the
above-mentioned symmetry reduction in quantum mechanics, we shall show that
the symmetry consideration of the two coupled oscillators leads to the
fifteen Dirac matrices.

In Sec. \ref{2couple}, we construct transformation matrices for the
coupled oscillator problem in classical mechanics.  In Sec. \ref{canno},
we show that the $Sp(4)$ symmetry is not enough for full understanding of
two coupled oscillators in classical mechanics, and that the group $SL(4,r)$
is needed. It is pointed out that $SL(4,r)$ transformations are not always
canonical, and we need non-canonical transformations to deal with the
classical oscillator problem.  In Sec. \ref{isomor}, we study in detail
the local isomorphism between the group $SL(4,r)$ and $O(3,3)$, and use this
isomorphism to construct the $Sp(4)$ subgroups of $SL(4,r)$.  It is like
constructing the $O(3,2)$-like subgroups of $O(3,3)$.
In Sec. \ref{qmco}, it is noted that there are three $O(3,2)$-like and
three $O(2,3)$-like subgroups in $O(3,3)$.  There are therefore six $Sp(4)$
subgroups in $SL(4,r)$.  It is then shown that one of them is canonical for
a given phase-space coordinate system, and the remaining five are not.

\section{Construction of the $Sp(4)$ Symmetry Group from Two Coupled
Oscillators}\label{2couple}

Let us consider a system of two coupled harmonic oscillators.  The
Hamiltonian for this system is
\begin{equation}\label{hamil0}
H = {1\over 2}\left\{{1\over m}_{1}p^{2}_{1} +
{1\over m}_{2}p^{2}_{2} +
A' x^{2}_{1} + B' x^{2}_{2} + C' x_{1} x_{2} \right\}.
\end{equation}
where
\begin{equation}
A' > 0, \qquad B' > 0, \qquad 4A'B' - C'^2 > 0 .
\end{equation}
By making scale changes of $x_{1}$ and $x_{2}$ to
$(m_{1}/m_{2})^{1/4} x_{1}$ and $(m_{2}/m_{1})^{1/4} x_{2}$ respectively,
it is possible to make a canonical transformation of the above Hamiltonian
to the form \cite{knp91,arav89}
\begin{equation}\label{hamil1}
H = {1\over 2m}\left\{p^{2}_{1} + p^{2}_{2} \right\} +
{1\over 2}\left\{A x_{1}^{2} + B x^{2}_{2} + C x_{1} x_{2} \right\} ,
\end{equation}
with $m = (m_{1}m_{2})^{1/2}.$  We can decouple this Hamiltonian
by making the coordinate transformation:
\begin{equation}\label{coordy}
\pmatrix{y_{1} \cr y_{2}} = \pmatrix{\cos (\alpha /2) & -\sin (\alpha /2)
\cr \sin (\alpha /2) & \cos (\alpha /2)} \pmatrix{x_{1} \cr x_{2}}.
\end{equation}
Under this rotation, the kinetic energy portion of the Hamiltonian in
Eq.(\ref{hamil1}) remains
invariant.  Thus we can achieve the decoupling by diagonalizing the
potential energy.  Indeed, the system becomes diagonal if the angle
$\alpha$ becomes
\begin{equation}\label{eq.34}
\tan \alpha  = {C\over B - A} .
\end{equation}
This diagonalization procedure is well known.

As we did in Ref. \cite{hkny93}, we introduce the new parameters $K$ and
$\eta$ defined as
\begin{equation}\label{eq.35}
K = \sqrt{AB - C^{2}/4} , \qquad  \exp {(- 2\eta)} =
\frac {A + B + \sqrt{(A - B)^{2} + C^{2}}}{\sqrt{4AB - C^{2}}} ,
\end{equation}
in addition to the rotation angle $\alpha$.  In terms of this
new set of variables, $A, B$ and $C$ take the form
\begin{eqnarray}\label{eq.36}
A&=&K \left(e^{2\eta} \cos^{2} {\alpha \over 2} +
e^{-2\eta} \sin^{2}{\alpha \over 2} \right) , \nonumber \\[3mm]
B&=&K \left(e^{2\eta} \sin^{2} {\alpha \over 2} +
e^{-2\eta} \cos^{2}{\alpha \over 2} \right) , \nonumber \\[3mm]
C&=&K \left(e^{-2\eta} - e^{2\eta} \right) \sin\alpha .
\end{eqnarray}
the Hamiltonian can be written as
\begin{equation}\label{hamil2}
H = {1\over 2m} \left\{q^{2}_{1} + q^{2}_{2} \right\} +
{K\over 2}\left\{e^{2\eta } y^{2}_{1} + e^{-2\eta } y^{2}_{2}\right\} ,
\end{equation}
where $y_{1}$ and $y_{2}$ are defined in Eq.(\ref{coordy}), and
\begin{equation}\label{eq.38}
\pmatrix{q_{1} \cr q_{2}} = \pmatrix{\cos (\alpha /2) & -\sin (\alpha /2)
\cr \sin (\alpha /2) & \cos (\alpha /2)} \pmatrix{p_{1} \cr p_{2}}.
\end{equation}
This form will be our starting point.  The above rotation together
with that of Eq.(\ref{coordy}) is generated by $S_{3}$.

If we measure the coordinate variable in units of $(mK)^{1/4}$, and use
$(mK)^{-1/4}$ for the momentum variables, the Hamiltonian takes the form
\begin{equation}\label{hamil3}
H = {1\over 2} e^{\eta}
\left(e^{-\eta}q^{2}_{1} + e^{\eta}y^{2}_{1} \right)
+ {1 \over 2} e^{-\eta}
\left(e^{\eta}q^{2}_{2} + e^{-\eta}y^{2}_{2} \right) ,
\end{equation}
where the Hamiltonian is measured in units of $\omega = \sqrt{K/m}$.
If $\eta = 0$, the system becomes decoupled, and the Hamiltonian becomes
\begin{equation}\label{hamil4}
H = {1\over 2} \left(p^{2}_{1} + x^{2}_{1}\right)
+ {1 \over 2} \left(p^{2}_{2} + x^{2}_{2} \right).
\end{equation}
In this paper, we are interested in the transformation of this decoupled
Hamiltonian into the most general form given in Eq.(\ref{hamil3}).

It is important to note that the Hamiltonian of Eq.(\ref{hamil4}) cannot be
obtained from Eq.(\ref{hamil3}) by a canonical transformation.	For this
reason, the Hamiltonian of the form
\begin{equation}\label{hamil5}
H' = {1\over 2}\left(e^{-\eta}q^{2}_{1} + e^{\eta}y^{2}_{1} \right)
+ {1\over 2}\left(e^{\eta}q^{2}_{2} + e^{-\eta}y^{2}_{2} \right)
\end{equation}
may play an important role in our discussion.  This Hamiltonian can be
transformed into the decoupled form of Eq.(\ref{hamil4}) through a
canonical transformation.  We will eventually have to face the problem of
transforming the above form to H of Eq.(\ref{hamil3}), and we shall do this
in Sec. \ref{canno}.

In this section, we are interested in transformations which will bring
the uncoupled Hamiltonian of Eq.(\ref{hamil4}) to $H'$.  For the two
uncoupled oscillators, we can start with the coordinate system:
\begin{equation}\label{coord1}
\left(\eta _{1}, \eta _{2}, \eta _{3}, \eta _{4} \right) = \left(x_{1},
p_{1}, x_{2}, p_{2} \right) .
\end{equation}
This coordinate system is different from the traditional coordinate system
where the coordinate variables are ordered as $\left(x_{1}, x_{2}, p_{1},
p_{2} \right)$.  This unconventional coordinate system does not change the
physics or mathematics of the problem, but is convenient for studying
the uncoupled system as well as expanding and shrinking phase spaces.

Since the two oscillators are independent, it is possible to
perform linear canonical transformations on each coordinate separately.
The canonical transformation in the first coordinate system is generated by
\begin{equation}\label{eq.21}
A_{1} = {1 \over 2} \pmatrix{\sigma _{2} & 0 \cr 0 & 0 } , \qquad
B_{1} = {i \over 2} \pmatrix{\sigma _{3} & 0 \cr 0 & 0 } , \qquad
C_{1} = {i \over 2} \pmatrix{\sigma _{1} & 0 \cr 0 & 0 } .
\end{equation}
These generators satisfy the Lie algebra:
\begin{equation}\label{commuo21}
\left[A_{1}, B_{1}\right] = iC_{1}, \qquad
\left[B_{1}, C_{1}\right] = -iA_{1}, \qquad
\left[C_{1}, A_{1}\right] = iB_{1}. \qquad
\end{equation}
It is also well known that this set of commutation relations is
identical to that for the $(2 + 1)$-dimensional Lorentz group.	Linear
canonical transformations on the second coordinate are generated by
\begin{equation}\label{eq.23}
A_{2} = {1 \over 2} \pmatrix{0 & 0 \cr 0 & \sigma _{2}} , \qquad
B_{2} = {i \over 2} \pmatrix{0 & 0 \cr 0 & \sigma _{3}} , \qquad
C_{2} = {i \over 2} \pmatrix{0 & 0 \cr 0 & \sigma _{1}} .
\end{equation}
These generators also satisfy the Lie algebra of
Eq.(\ref{commuo21}).  We are interested here in constructing the symmetry
group for the coupled oscillators by soldering two $Sp(2)$ groups generated
by $A_{1}, B_{1}, C_{1}$ and $A_{2}, B_{2}, C_{2}$ respectively.

It will be more convenient to use the linear combinations:
$$
A_{+} = A_{1} + A_{2}, \qquad B_{+} = B_{1} + B_{2}, \qquad
C_{+} = C_{1} + C_{2},
$$
\begin{equation}\label{eq.24}
A_{-} = A_{1} - A_{2}, \qquad B_{-} = B_{1} - B_{2}, \qquad
C_{-} = C_{1} - C_{2},
\end{equation}
These matrices take the form

$$
A_{+} = {1 \over 2}\pmatrix{\sigma _{2} & 0 \cr 0 & \sigma _{2}} , \qquad
B_{+} = {i \over 2}\pmatrix{\sigma _{3} & 0 \cr 0 & \sigma_{3}}, \qquad
C_{+} = {i \over 2}\pmatrix{\sigma _{1} & 0 \cr 0 & \sigma _{1}},
$$

\begin{equation}\label{eq.25}
A_{-} = {1 \over 2}\pmatrix{\sigma _{2} & 0 \cr 0 & -\sigma _{2}}, \qquad
B_{-} = {i \over 2}\pmatrix{\sigma _{3} & 0 \cr 0 & -\sigma _{3}}, \qquad
C_{-} = {i \over 2}\pmatrix{\sigma _{1} & 0 \cr 0 & -\sigma _{1}}.
\end{equation}
The sets $\left(A_{+}, B_{+}, C_{+}\right)$ and
$\left(A_{+}, B_{-}, C_{-}\right)$ satisfy the Lie algebra of
Eq.(\ref{commuo21}).  The same is true for
$\left(A_{-}, B_{+}, C_{-}\right)$ and $\left(A_{-}, B_{-}, C_{+}\right)$.

Next, let us couple the oscillators through a rotation generated by
\begin{equation}\label{A0}
A_{0} = {i \over 2}\pmatrix{0 & -I \cr I & 0} .
\end{equation}
In view of the fact that the first two coordinate variables are for the
phase space of the first oscillator, and the third and fourth are for the
second oscillator, this matrix generates parallel rotations in the
$(x_{1},x_{2})$
and $(p_{1},p_{2})$ coordinates.  As the coordinates $(x_{1},x_{2})$ are
coupled through a two-by-two matrix, the coordinate $(p_{1},p_{2})$ are
coupled through the same two-by-two matrix.

Then, $A_{0}$ commutes with $A_{+}, B_{+}, C_{+}$, and the following
commutation relations generate new operators $A_{3}, B_{3}$ and
$C_{3}$:
\begin{equation}\label{eq.27}
\left[A_{0}, A_{-}\right] = iA_{3}, \qquad
\left[A_{0}, B_{-}\right] = iB_{3}, \qquad
\left[A_{0}, C_{-}\right] = iC_{3}, \qquad
\end{equation}
where
\begin{equation}\label{eq.28}
A_{3} = {1 \over 2}\pmatrix{0 & \sigma _{2} \cr \sigma _{2} & 0} , \qquad
B_{3} = {i \over 2}\pmatrix{0 & \sigma _{3} \cr \sigma _{3} & 0} , \qquad
C_{3} = {i \over 2}\pmatrix{0 & \sigma _{1} \cr \sigma _{1} & 0} .
\end{equation}

In this section, we started with the generators of the symmetry groups for
two independent oscillators.  They are $A_{1}, B_{1}, C_{1}$ and
$A_{2}, B_{2}, C_{2}$.	We then introduced $A_{0}$ which generates coupling
of two oscillators.  This processes produced three additional generators
$A_{3}, B_{3}, C_{3}$.	It is remarkable that $C_{3}, B_{3}$ and $A_{+}$
form the set of generators for another $Sp(2)$ group.  They satisfy the
commutation relations
\begin{equation}
[B_{3}, C_{3}] = -iA_{+} , \qquad  [C_{3}, A_{+}] = iB_{3} , \qquad
[A_{+}, B_{3}] = iC_{3} .
\end{equation}
The same can be said about the sets $A_{+}, B_{1}, C_{1}$ and
$A_{+}, B_{2}, C_{2}$.	These Sp(2)-like groups are associated with the
coupling of the two oscillators.

\section{Canonical and Non-canonical Transformations in Classical
Mechanics}\label{canno}

For a dynamical system consisting of two pairs of canonical variables
$x_{1}, p_{1}$ and $x_{2}, p_{2}$, we have introduced the coordinate system
$\left(\eta _{1}, \eta _{2}, \eta _{3}, \eta _{4} \right)$ defined in
Eq.(\ref{coord1}).

The transformation of the variables from $\eta _{i}$ to $\xi _{i}$ is
canonical if
\begin{equation}\label{symp}
M J \tilde{M} = J ,
\end{equation}
where
$$
M_{ij} = {\partial \over \partial \eta _{j}}\xi _{i},
$$
\noindent and
\begin{equation}\label{JKQ}
J = \pmatrix{0 & 1 & 0 & 0 \cr -1 & 0 & 0 & 0
\cr 0 & 0 & 0 & 1 \cr 0 & 0 & -1 & 0} .
\end{equation}
This form of the J matrix appears different from the traditional literature,
because we are using the new coordinate system.  In order to avoid possible
confusion and to maintain continuity with our earlier publications, we give
in the Appendix the expressions for the J matrix and the ten generators of
the $Sp(4)$ group in the traditional coordinate system.  There are four
rotation generators and six squeeze generators in this group.

In this new coordinate system, the rotation generators take the form
$$
L_{1} = {-1\over 2}\pmatrix{0&\sigma _{2} \cr \sigma _{2} & 0}, \qquad
L_{2} = {i\over 2}\pmatrix{ 0 & -I \cr I & 0}  ,
$$
\begin{equation}\label{rotKQ}
L_{3} = {-1\over 2}\pmatrix{\sigma _{2}&0 \cr 0 &-\sigma _{2}}, \qquad
S_{3} = {1\over 2}\pmatrix{\sigma _{2} & 0 \cr 0 & \sigma _{2} } .
\end{equation}
The squeeze generators become
$$
K_{1} = {i\over 2}\pmatrix{\sigma _{1} & 0 \cr 0 & -\sigma _{1} },
\qquad
K_{2} = {i\over 2}\pmatrix{\sigma _{3} & 0 \cr 0 & \sigma _{3}} , \qquad
K_{3} = -{i\over 2}\pmatrix{0 & \sigma _{1} \cr \sigma _{1} & 0} ,
$$

\begin{equation}\label{sqKQ}
Q_{1} = {i\over 2}\pmatrix{-\sigma _{3} & 0 \cr 0 & \sigma _{3}}, \qquad
Q_{2} = {i\over 2}\pmatrix{\sigma _{1} & 0 \cr 0 & \sigma _{1}} , \qquad
Q_{3} = {i\over 2}\pmatrix{0 & \sigma _{3} \cr \sigma _{3} & 0} .
\end{equation}

There are now ten generators.  They form the Lie algebra for the $Sp(4)$
group:
$$
[L_{i}, L_{j}] = i\epsilon _{ijk} L_{k}, \qquad
[L_{i}, S_{3}] = 0, \qquad
$$
$$
[L_{i}, K_{j}] = i\epsilon _{ijk} K_{k}, \qquad
[L_{i}, Q_{j}] = i\epsilon _{ijk} Q_{k}, \qquad
$$
$$
[K_{i}, K_{j}] = [Q_{i}, Q_{j}] = -i\epsilon _{ijk} L_{k} , \qquad
[K_{i}, Q_{j}] = -i\delta _{ij} S_{3} ,
$$
\begin{equation}\label{LieSp4}
[K_{i}, S_{3}] = -iQ_{i} , \qquad [Q_{i}, S_{3}] = iK_{i} .
\end{equation}

Indeed, these matrices can be identified with the ten matrices introduced
in Sec.\ref{2couple} in the following manner.
$$
A_{+} = S_{3}, \quad A_{-} = -L_{3}, \quad A_{3} = -L_{1} , \quad
A_{0} = L_{2},
$$
$$
B_{+} = K_{2}, \qquad B_{-} = -Q_{1}, \qquad B_{3} = Q_{3} , \\
$$
\begin{equation}
C_{+} = Q_{2}, \qquad C_{-} = K_{1}, \qquad C_{3} = -K_{3} .
\end{equation}

In Sec. \ref{2couple}, we started with the $Sp(2)$ symmetry for each of
the oscillator, and introduced the parallel rotation to couple the system.
It is interesting to note that this process leads to the $Sp(4)$ symmetry.

We have chosen the non-traditional phase space coordinate system given in
Eq.(\ref{coord1}) in order to study the coupling more effectively.   The
$A_{0}$ matrix given in Eq.(\ref{A0}) generates the coupling of two phase
spaces by rotation.  Within this coordinate system, we are interested in
relative adjustments of the sizes of the two phase spaces.  Indeed, in
order to transform the Hamiltonian of Eq.(\ref{hamil5}) to that of
Eq.(\ref{hamil3}), we have to expand one phase space while contracting
the other.  For this purpose, we need the generators of the form
\begin{equation}\label{G3}
G_{3} = {i\over 2} \pmatrix{I & 0 \cr 0 & -I} .
\end{equation}
This matrix generates scale transformations in phase space.  The
transformation leads to a radial expansion of the phase space of the
first coordinate \cite{kili89} and contracts the phase space of the
second coordinate.  What is the physical significance of this operation?
The expansion of phase space leads to an increase in uncertainty and
entropy.  Mathematically speaking, the contraction of the
second coordinate should cause a decrease in uncertainty and entropy.
Can this happen?  The answer is clearly No, because it will violate the
uncertainty principle.	This question will be addressed in future
publications.

In the meantime, let us study what happens when the matrix $G_{3}$ is
introduced into the set of matrices given in Eq.(\ref{rotKQ}) and
Eq.(\ref{sqKQ}).  It commutes with $S_{3}, L_{3}, K_{1}, K_{2}, Q_{1}$,
and $Q_{2}$.  However, its commutators with the rest of the matrices
produce four more generators:
\begin{equation}
\left[G_{3}, L_{1}\right] = iG_{2} , \qquad
\left[G_{3}, L_{2}\right] = -iG_{1} , \qquad
\left[G_{3}, K_{3}\right] = iS_{2} , \qquad
\left[G_{3}, Q_{3}\right] = -iS_{1} ,
\end{equation}
with
\begin{eqnarray}
G_{1} &=& {i\over 2}\pmatrix{0 & I \cr I & 0} , \qquad
G_{2} = {1\over 2}\pmatrix{0 & -\sigma_{2} \cr
\sigma_{2} & 0} ,  \nonumber \\[3mm]
S_{1} &=& {-i\over 2}\pmatrix{0 & -\sigma_{3} \cr \sigma_{3} & 0} , \qquad
S_{2} = {i\over 2}\pmatrix{0 & -\sigma_{1} \cr \sigma_{1} & 0} .
\end{eqnarray}
If we take into account the above five generators in addition to the
ten generators of $Sp(4)$, there are fifteen generators.  These generators
satisfy the following set of commutation relations.
$$
[L_{i}, L_{j}] = i\epsilon _{ijk} L_{k}, \qquad
[S_{i}, S_{j}] = i\epsilon _{ijk} S_{k}, \qquad
[L_{i}, S_{j}] = 0, \qquad
$$
$$
[L_{i}, K_{j}] = i\epsilon _{ijk} K_{k}, \qquad
[L_{i}, Q_{j}] = i\epsilon _{ijk} Q_{k}, \qquad
[L_{i}, G_{j}] = i\epsilon _{ijk} G_{k},
$$
$$
[K_{i}, K_{j}] = [Q_{i}, Q_{j}] = [Q_{i}, Q_{j}]
= -i\epsilon _{ijk} L_{k} , \qquad
$$
$$
[K_{i}, Q_{j}] = -i\delta _{ij} S_{3} , \qquad
[Q_{i}, G_{j}] = -i\delta _{ij} S_{1} , \qquad
[G_{i}, K_{j}] = -i\delta _{ij} S_{2} .
$$
$$
[K_{i}, S_{3}] = -iQ_{i} , \qquad [Q_{i}, S_{3}] = iK_{i} ,\qquad
[G_{i}, S_{3}] = 0 ,
$$
$$
[K_{i}, S_{1}] = 0 , \qquad [Q_{i}, S_{1}] = -iG_{i} ,\qquad
[G_{i}, S_{1}] = iQ_{i} ,
$$

\begin{equation}
[K_{i}, S_{2}] = iG_{i} , \qquad [Q_{i}, S_{2}] = 0 ,\qquad
[G_{i}, S_{2}] = -iK_{i} .
\end{equation}

Indeed, the ten $Sp(4)$ generators together with the five new generators
form the Lie algebra for the group $SL(4,r)$.  This group is known to be
locally isomorphic to the Lorentz group $O(3,3)$ with three space variables
and three time variables.  This means that we can study the symmetry of
the two coupled oscillators with this high-dimensional Lorentz group.  It
is also known that the fifteen Dirac matrices in the Majorana form can
serve as the generators of $SL(4,r)$.  Thus, the coupled oscillator can
also serve as a model for the Dirac matrices.  Indeed, this
higher-dimensional group could lead to a much richer picture of symmetry
than is known today.  In the meantime, let us study the local isomorphism
between $O(3,3)$ and $SL(4,r)$.

\section{Local Isomorphism between $O(3,3)$ and SL(4,r)}\label{isomor}

In Secs. \ref{2couple} and \ref{canno}, we constructed the fifteen
generators of the group $SL(4,r)$ from the coupled oscillator system.
In this section, we write down the generators of the $O(3,3)$ group
and confirm that $O(3,3)$ is locally isomorphic to $SL(4,r)$.  One
immediate advantage is to use this isomorphism to construct
$Sp(4)$-like subgroups of $SL(4,r)$.

In the Lorentz group $O(3,1)$, there are three rotation generators and
three boost generators.  In $O(3,2)$, there are three rotation generators
for the three space-like coordinates, and one rotation generator for the
two time-like coordinates.  We can make boosts along the three different
space-like directions with respect to each time-like variable.  There are
therefore two sets of three boost generators for this system with two
time-like variables.  In this manner, we have studied effectively the
$Sp(2)$ subgroups of the group $Sp(4)$.  It was interesting to note that
there are three $O(1,2)$ subgroups with one space-like and two time-like
variables.  This is translated into the three corresponding $Sp(2)$
subgroups in $Sp(4)$.  The algebraic property of these $O(1,2)$ like
subgroups is the same as those for $O(2,1)$-like subgroups.

In the present case of $O(3,3)$, there are now three time-like coordinates.
For this three-dimensional space, there are three rotation generators, and
there are three sets of boost generators.  We should not forget the three
rotation generators operating in the three-dimensional space-like space.
This is one convenient way to classify the six rotation and nine boost
generators in the $O(3,3)$ as well as in the $SL(4,r)$ group where the
boost operators become squeeze operators.

With these points in mind, we introduce a six-dimensional space with three
space-like variables $x, y, z$ and three time-like variables $s, t, u$.
These variables can be ordered as $(x, y, z, s, t, u)$, and transformations
can be performed by six-by-six matrices.

Let us start with the $O(3,2)$ subgroup which was discussed in our earlier
papers.  The transformations operate in the five-dimensional subspace
$(x, y, z, s, t)$.  We can still write six-by-six matrices for the
generators of this group with zero elements on both sixth rows and columns.
Then according to Ref. \cite{hkny93}, the generators take the form

\begin{eqnarray}\label{rotL}
L_{1} &=& \pmatrix{0&0&0&0&0&0 \cr 0&0&-i&0&0&0 \cr 0&i&0&0&0&0
\cr 0&0&0&0&0&0 \cr 0&0&0&0&0&0 \cr 0&0&0&0&0&0} , \hspace{6mm}
L_{2} = \pmatrix{0&0&i&0&0&0 \cr 0&0&0&0&0&0 \cr -i&0&0&0&0&0
\cr 0&0&0&0&0&0 \cr 0&0&0&0&0&0 \cr 0&0&0&0&0&0} , \nonumber \\[3mm]
L_{3} &=& \pmatrix{0&-i&0&0&0&0 \cr i&0&0&0&0&0 \cr 0&0&0&0&0&0
\cr 0&0&0&0&0&0 \cr 0&0&0&0&0&0 \cr 0&0&0&0&0&0} .
\end{eqnarray}

The Lorentz boosts in the subspace of $(x, y, z, t)$ are generated by

\begin{eqnarray}\label{boostK}
K_{1} &=& \pmatrix{0&0&0&i&0&0 \cr 0&0&0&0&0&0 \cr 0&0&0&0&0&0
\cr i&0&0&0&0&0 \cr 0&0&0&0&0&0 \cr 0&0&0&0&0&0} , \hspace{6mm}
K_{2} = \pmatrix{0&0&0&0&0&0 \cr 0&0&0&i&0&0 \cr 0&0&0&0&0&0
\cr 0&i&0&0&0&0 \cr 0&0&0&0&0&0 \cr 0&0&0&0&0&0} , \nonumber \\[3mm]
K_{3} &=& \pmatrix{0&0&0&0&0&0 \cr 0&0&0&0&0&0 \cr 0&0&0&i&0&0
\cr 0&0&i&0&0&0 \cr 0&0&0&0&0&0 \cr 0&0&0&0&0&0} .
\end{eqnarray}

These three boost generators, together with the rotation generators of
Eq.(\ref{rotL}), form the Lie algebra for the Lorentz group
applicable to the four-dimensional Minkowski space of $(x, y, z, t)$.
The same is true for the space of $(x, y, z, s)$ with the boost
generators:
\begin{eqnarray}\label{boostQ}
Q_{1} &=& \pmatrix{0&0&0&0&i&0 \cr 0&0&0&0&0&0 \cr 0&0&0&0&0&0
\cr 0&0&0&0&0&0 \cr i&0&0&0&0&0 \cr 0&0&0&0&0&0} , \hspace{6mm}
Q_{2} = \pmatrix{0&0&0&0&0&0 \cr 0&0&0&0&i&0 \cr 0&0&0&0&0&0
\cr 0&0&0&0&0&0 \cr 0&i&0&0&0&0 \cr 0&0&0&0&0&0} , \nonumber \\[3mm]
Q_{3} &=& \pmatrix{0&0&0&0&0&0 \cr 0&0&0&0&0&0 \cr 0&0&0&0&i&0
\cr 0&0&0&0&0&0 \cr 0&0&i&0&0&0 \cr 0&0&0&0&0&0} .
\end{eqnarray}
The above two Lorentz groups have nine generators.  If we attempt
to form a closed set of commutation relations, we end up with an
additional
\begin{equation}\label{rotS3}
S_{3} = \pmatrix{0&0&0&0&0&0 \cr 0&0&0&0&0&0 \cr 0&0&0&0&0&0
\cr 0&0&0&0&-i&0 \cr 0&0&0&i&0&0 \cr 0&0&0&0&0&0} ,
\end{equation}
which will generate rotations in the two-dimensional space of $s$ and
$t$.  These ten generators form a closed set of commutations relations.

According to the commutation relations given in Sec. \ref{canno}, we can
construct two more rotation generators:
\begin{equation}\label{rotS12}
S_{1} = \pmatrix{0&0&0&0&0&0 \cr 0&0&0&0&0&0 \cr 0&0&0&0&0&0
\cr 0&0&0&0&0&0 \cr 0&0&0&0&0&-i \cr 0&0&0&0&i&0} , \hspace{6mm}
S_{2} = \pmatrix{0&0&0&0&0&0 \cr 0&0&0&0&0&0 \cr 0&0&0&0&0&0
\cr 0&0&0&0&0&i \cr 0&0&0&0&0&0 \cr 0&0&0&-i&0&0} ,
\end{equation}
which, together with $S_{3}$, satisfy the Lie algebra for the
three-dimensional rotation group.  In addition, there are three additional
boost generators:
\begin{eqnarray}\label{boostG}
G_{1} &=& \pmatrix{0&0&0&0&0&i \cr 0&0&0&0&0&0 \cr 0&0&0&0&0&0
\cr 0&0&0&0&0&0 \cr 0&0&0&0&0&0 \cr i&0&0&0&0&0} , \hspace{6mm}
G_{2} = \pmatrix{0&0&0&0&0&0 \cr 0&0&0&0&0&i \cr 0&0&0&0&0&0
\cr 0&0&0&0&0&0 \cr 0&0&0&0&0&0 \cr 0&i&0&0&0&0} , \nonumber \\[3mm]
G_{3} &=& \pmatrix{0&0&0&0&0&0 \cr 0&0&0&0&0&0 \cr 0&0&0&0&0&i
\cr 0&0&0&0&0&0 \cr 0&0&0&0&0&0 \cr 0&0&i&0&0&0} .
\end{eqnarray}

These generators satisfy the commutation relations given in
Sec. \ref{canno}, and this confirms the local isomorphism between
$O(3,3)$ and $SL(4,r)$ which we constructed in this paper from the coupled
oscillators.

Let us now go back to the $J$ matrix of Eq.(\ref{JKQ}).  This $J$ matrix
is proportional to $S_{3}$ which is one of the rotation operators
applicable to three time-like coordinates.  This means that $S_{1}$ and
$S_{2}$ can also play their respective roles as the $J$ matrix.  If
$S_{1}$ is proportional to the J matrix, the $O(3,2)$ subgroup operates
in the five dimensional space of $(x, y, z, t, u)$.  If we use $S_{2}$,
the subspace is $(x, y, z, s, u)$.

Because there are three time-like coordinates, there are three $O(2,3)$
subgroups.  In these cases, each one of the rotation generators
$L_{1}, L_{2}$, and $L_{3}$ can serve as the J matrix.  If $L_{2}$ is
chosen, for instance, the five-dimensional subspace is $(x, z, s, t, u)$.
The corresponding $Sp(4)$-like subgroup is discussed in detail in the
Appendix.

It has been known for sometime that the fifteen Dirac matrices can serve
as the generators of the group locally isomorphic to $O(3,3)$ \cite{dir63}.
In his recent paper \cite{lee95}, Lee showed that the generators of
$SL(4,r)$ indeed constitute the Dirac matrices in the Majorana
representation.  Thus, we have shown in this paper that the system of two
coupled oscillators can serve as a mechanical model for the Dirac matrices.

\section{Quantum Mechanics of Coupled Oscillators}\label{qmco}
We had to construct the $O(3,3)$-like symmetry group in order to
transform the Hamiltonian of two uncoupled identical oscillators given
in Eq.(\ref{hamil4}) to the Hamiltonian $H'$ of Eq.(\ref{hamil5}) and
then into the Hamiltonian $H$ of Eq.(\ref{hamil3}).  The ground-state
wave function for the two uncoupled oscillators takes the form
\begin{equation}\label{wf1}
\psi_{0} (x_{1},x_{2}) = \frac {1}{\sqrt{\pi}}
\exp{ \left\{- {1\over 2}(x^{2}_{1} + x^{2}_{2})
\right\} } .
\end{equation}
We do not have to explain in this paper how to construct wave functions
for excited states.  Our problem is how to transform the above wave function
into
\begin{equation}\label{wf2}
\psi_{0} (x_{1},x_{2}) = \frac {1}{\sqrt{\pi}}
\exp{ \left\{- {1\over 2}(e^{\eta } y^{2}_{1} + e^{-\eta } y^{2}_{2})
\right\} } ,
\end{equation}
which is the ground-state wave function for the Hamiltonian $H'$ of
Eq.(\ref{hamil5}).  In terms of the $x_{1}$ and $x_{2}$ variables, this
wave function can be written as
\begin{eqnarray}\label{wf3}
\lefteqn{\psi (x_{1},x_{2}) = \frac{1}{\sqrt{\pi}} \exp \left\{- {1\over 2}
\left [ e^{\eta }(x_{1} \cos {\alpha \over 2} - x_{2} \sin {\alpha
\over 2})^{2} \right. \right. } \nonumber \\ \vspace*{.5ex} \nonumber \\
\mbox{ } & \mbox{ } & \mbox{ } \hspace{4cm} \left.  \left. + e^{-\eta
}(x_{1} \sin {\alpha\over 2} + x_{2} \cos {\alpha \over 2})^{2} \right ]
\right\} .\hspace*{2cm}
\end{eqnarray}
It has been shown that there exists a unitary transformation which changes
the ground-state wave function of Eq.(\ref{wf1}) to the above
form~\cite{hkn90}.  In general, unitary transformations are generated by
the differential operators \cite{dir63}:
\begin{eqnarray}\label{hatdag}
\hat{L}_{1} &=& {1\over 2}\left(a^{\dag }_{1}a_{2} + a^{\dag }_{2}a_{1}
\right) ,\qquad \hat{L}_{2} = {1\over 2i}\left(a^{\dag }_{1}a_{2} -
a^{\dag }_{2}a_{1}\right) ,  \nonumber \\[3mm]
\hat{L}_{3} &=& {1\over 2}\left(a^{\dag}_{1}a_{1} -
a^{\dag}_{2}a_{2} \right) , \qquad
\hat{S}_{3} = {1\over 2}\left(a^{\dag}_{1}a_{1} +
a_{2}a^{\dag}_{2} \right) ,   \nonumber \\[3mm]
\hat{K}_{1} &=& -{1\over 4}\left(a^{\dag}_{1}a^{\dag}_{1} + a_{1}a_{1} -
a^{\dag}_{2}a^{\dag}_{2} - a_{2}a_{2}\right) ,	 \nonumber \\[3mm]
\hat{K}_{2} &=& {i\over 4}\left(a^{\dag}_{1}a^{\dag}_{1} - a_{1}a_{1} +
a^{\dag}_{2}a^{\dag}_{2} - a_{2}a_{2}\right) ,	 \nonumber \\[3mm]
\hat{K}_{3} &=& {1\over 2}\left(a^{\dag}_{1}a^{\dag}_{2} +
a_{1}a_{2}\right) ,   \nonumber \\[3mm]
\hat{Q}_{1} &=& -{i\over 4}\left(a^{\dag}_{1}a^{\dag}_{1} - a_{1}a_{1} -
a^{\dag}_{2}a^{\dag}_{2} + a_{2}a_{2} \right) ,   \nonumber \\[3mm]
\hat{Q}_{2} &=& -{1\over 4}\left(a^{\dag}_{1}a^{\dag}_{1} + a_{1}a_{1} +
a^{\dag}_{2}a^{\dag}_{2} + a_{2}a_{2} \right) ,  \nonumber \\[3mm]
\hat{Q}_{3} &=& {i\over 2}\left(a^{\dag}_{1}a^{\dag}_{2} -
a_{1}a_{2} \right) .
\end{eqnarray}
where $a^{\dag}$ and a are the step-up and step-down operators applicable
to harmonic oscillator wave functions.  These operators satisfy the Lie
algebra for the $Sp(4)$ group in Eq.(\ref{LieSp4}), and there is a
one-to-one correspondence between the hatted operators in this section and
the unhatted operators in Sec. \ref{canno}.

Next, we are led to the question of whether there is a transformation in
quantum mechanics which corresponds to the transformation of $H'$ of
Eq.(\ref{hamil5}) into $H$ of Eq.(\ref{hamil3}).  It was noted in
Sec.~\ref{canno} that this transformation is non-canonical.  At the
present time, we do not know how to translate non-canonical transformations
into the language of the Schr\"odinger picture of quantum mechanics where
only unitary transformations are allowed.  We are thus unable add five
more hatted operators to the above list of ten generators.

Under these circumstances, the best we can do is to use the same wave
function for both $H$ and $H'$ with different expressions for the
eigenvalues.  For $H$, the energy eigenvalues are
\begin{equation}
E_{n_{1},n_{2}} = n_{1} + n_{2} + 1 ,
\end{equation}
where $n_{1}$ and $n_{2}$ are the excitation numbers for the first and
second modes respectively.  On the other hand, the eigenvalues for $H'$ are
\begin{equation}
E_{n_{1},n_{2}} = e^{\eta}n_{1} + e^{-\eta}n_{2} + 1 .
\end{equation}

On the other hand, there is a provision for nonunitary transformations in
the density matrix formulation of quantum mechanics \cite{fey72,neu32}.
One way to deal with this problem for the present case is to use the Wigner
phase space picture of quantum mechanics \cite{wig32}.

For two-mode problems, the Wigner function is defined as \cite{knp91}
\begin{eqnarray}\label{wigf1}
\lefteqn{W(x_{1},x_{2}; p_{1},p_{2}) = \left({1\over \pi} \right)^{2}
\int \exp \left\{- 2i (p_{1}y_{1} + p_{2}y_{2}) \right\} } \nonumber\\[3mm]
\mbox{ } & \mbox{ } & \mbox{ }
\times \psi^{*}(x_{1} + y_{1}, x_{2} + y_{2})
\psi (x_{1} - y_{1}, x_{2} - y_{2}) dy_{1} dy_{2} .\hspace*{2cm}
\end{eqnarray}
The Wigner function corresponding to the oscillator wave function of
Eq.(\ref{wf2}) is
\begin{eqnarray}\label{wigf2}
\lefteqn{ W(x_{1},x_{2};p_{1},p_{2}) = \left (\frac{1}{\pi} \right)^{2}
\exp \left\{-e^{\eta }(x_{1} \cos {\alpha \over 2}
- x_{2} \sin {\alpha \over 2})^{2} \right. }\nonumber \\ [3mm]
\mbox{ } & \mbox{ } & \mbox{ } \left.-
e^{-\eta }(x_{1} \sin {\alpha \over 2} +
x_{2} \cos {\alpha \over 2})^{2} - e^{-\eta }(p_{1} \cos {\alpha
\over 2} - p_{2} \sin {\alpha \over 2})^{2} \right. \nonumber \\[3mm]
\mbox{ } & \mbox { } & \mbox{ } \hspace{4cm} \left.
- e^{\eta }(p_{1} \sin {\alpha \over 2} + p_{2} \cos
{\alpha \over 2})^{2}\right\} .
\end{eqnarray}

Indeed, the Wigner function is defined over the four-dimensional phase
space of $(x_{1}, p_{1}, x_{2}, p_{2})$ just as in the case of classical
mechanics.  The unitary transformations generated by the operators of
Eq.(\ref{hatdag}) are translated into linear canonical transformations
of the Wigner function~\cite{hkn90}.  The canonical transformations are
generated by the differential operators~\cite{knp91}:
\begin{eqnarray}\label{rotphase}
L_{1} &=& +{i\over 2}\left\{\left(x_{1}{\partial \over \partial p_{2}} -
p_{2}{\partial \over \partial x_{1}} \right) +
\left(x_{2}{\partial \over \partial p_{1}} -
p_{1}{\partial \over \partial x_{2}} \right)\right\}, \nonumber \\[3mm]
L_{2} &=& -{i\over 2}\left\{\left(x_{1}{\partial \over \partial x_{2}} -
x_{2}{\partial \over \partial x_{1}}\right) +
\left(p_{1} {\partial \over \partial p_{2}} -
p_{2}{\partial \over \partial p_{1}}\right)\right\} ,\nonumber \\[3mm]
L_{3} &=& +{i\over 2}\left\{\left(x_{1}{\partial \over \partial p_{1}} -
p_{1}{\partial \over \partial x_{1}}\right) -
\left(x_{2}{\partial \over \partial p_{2}} -
p_{2}{\partial \over \partial x_{2}}\right)\right\} , \nonumber \\[3mm]
S_{3} &=& -{i\over 2}\left\{\left(x_{1}{\partial \over \partial p_{1}} -
p_{1}{\partial \over \partial x_{1}}\right) +
\left(x_{2}{\partial \over \partial p_{2}} -
p_{2}{\partial \over \partial x_{2}}\right)\right\} ,
\end{eqnarray}
and
\begin{eqnarray}\label{sqphase}
K_{1} &=& -{i\over 2}\left\{\left( x_{1}{\partial \over \partial p_{1}} +
p_{1}{\partial \over \partial x_{1}} \right) -
\left(x_{2}{\partial \over \partial p_{2}} +
p_{2}{\partial \over \partial x_{2}} \right)\right\}, \nonumber \\[3mm]
K_{2} &=& -{i\over 2}\left\{\left(x_{1}{\partial \over \partial x_{1}} -
p_{1}{\partial \over \partial p_{1}}\right) +
\left(x_{2}{\partial \over \partial x_{2}} -
p_{2}{\partial \over \partial p_{2}}\right)\right\} , \nonumber \\[3mm]
K_{3} &=& +{i\over 2}\left\{\left(x_{1}{\partial \over \partial p_{2}} +
p_{2}{\partial \over \partial x_{1}}\right) +
\left(x_{2}{\partial \over \partial p_{1}} +
p_{1}{\partial \over \partial x_{2}}\right)\right\} , \nonumber \\[3mm]
Q_{1} &=& +{i\over 2}\left\{\left(x_{1}{\partial \over \partial x_{1}} -
p_{1}{\partial \over \partial p_{1}}\right) -
\left(x_{2}{\partial \over \partial x_{2}} -
p_{2}{\partial \over \partial p_{2}}\right)\right\} ,\nonumber \\[3mm]
Q_{2} &=& -{i\over 2}\left\{\left(x_{1}{\partial \over \partial p_{1}} +
p_{1}{\partial \over \partial x_{1}}\right) +
\left(x_{2}{\partial \over \partial p_{2}} +
p_{2}{\partial \over \partial x_{2}}\right)\right\} ,\nonumber \\[3mm]
Q_{3} &=& -{i\over 2}\left\{\left(x_{2}{\partial \over \partial x_{1}} +
x_{1}{\partial \over \partial x_{2}} \right) -
\left(p_{2}{\partial \over \partial p_{1}} + p_{1}{\partial \over
\partial p_{2}}\right)\right\} .
\end{eqnarray}
These differential operators are the same as their matrix counterparts
given in Sec. \ref{canno}.

Unlike the case of the Schr\"odinger picture, it is possible to add five
noncanonical generators to the above list.  They are
\begin{eqnarray}\label{rotphase2}
S_{1} &=& +{i\over 2}\left\{\left(x_{1}{\partial \over \partial x_{2}} -
x_{2}{\partial \over \partial x_{1}} \right) -
\left(p_{1}{\partial \over \partial p_{2}} -
p_{2}{\partial \over \partial p_{1}} \right)\right\}, \nonumber \\[3mm]
S_{2} &=& -{i\over 2}\left\{\left(x_{1}{\partial \over \partial p_{2}} -
p_{2}{\partial \over \partial x_{1}}\right) +
\left(x_{2} {\partial \over \partial p_{1}} -
p_{1}{\partial \over \partial x_{2}}\right)\right\} ,
\end{eqnarray}
as well as three additional squeeze operators:

\begin{eqnarray}\label{sqphase2}
G_{1} &=& -{i\over 2}\left\{\left( x_{1}{\partial \over \partial x_{2}} +
x_{2}{\partial \over \partial x_{1}} \right) +
\left(p_{1}{\partial \over \partial p_{2}} +
p_{2}{\partial \over \partial p_{1}} \right)\right\}, \nonumber \\[3mm]
G_{2} &=& {i\over 2}\left\{\left(x_{1}{\partial \over \partial p_{2}} +
p_{2}{\partial \over \partial x_{1}}\right) -
\left(x_{2}{\partial \over \partial p_{1}} +
p_{1}{\partial \over \partial x_{2}}\right)\right\} , \nonumber \\[3mm]
G_{3} &=& -{i\over 2}\left\{\left(x_{1}{\partial \over \partial x_{1}} +
p_{1}{\partial \over \partial p_{1}}\right) +
\left(x_{2}{\partial \over \partial p_{1}} +
p_{1}{\partial \over \partial x_{2}}\right)\right\}.
\end{eqnarray}

These five generators perform well defined operations on the Wigner
function from the mathematical point of view.  However, the question is
whether these additional generators are acceptable in the present form of
quantum mechanics.

In order to answer this question, let us note that the uncertainty
principle in the phase-space picture of quantum mechanics is stated in
terms of the minimum area in phase space for a given pair of conjugate
variables.  The minimum area is determined by Planck's constant.  Thus we
are allowed to expand the phase space, but are not allowed to contract it.
With this point in mind, let us go back to $G_{3}$ of Eq.(\ref{G3}), which
generates transformations which simultaneously expand one phase space and
contract the other.  Thus, the $G_{3}$ generator is not acceptable in
quantum mechanics even though it generates well-defined mathematical
transformations of the Wigner function.

Unlike the matrix generators, the form of differential operators applicable
to the four-dimensional phase space is invariant under reordering of the
coordinate variables.  Of course, there are six different ways to choose
ten generators from fifteen to construct the $O(3,2)$-like subgroups.  Thus
However, only one of them can be accommodated in the present form of
quantum mechanics.  The rotation generators $S_{1}, S_{2}$ and the squeeze
generators $G_{1}, G_{2}, G_{3}$ cannot generate meaningful transformations
in quantum mechanics.  The question of whether they are useful in the
quantum world remains as an interesting future problem.

In this paper, we started with one of the most elementary physical systems,
and noted that it leads to one of the most sophisticated symmetry problems
in physics.  It is remarkable that the differential operators given in this
Section can serve as Dirac matrices.  These operators generate geometric
transformations in a four-dimensional space.  Indeed many attempts have
been made in the past to give geometric interpretations of
the Dirac matrices, and the present paper is not likely to be the last one.

\begin{appendix}

\section*{Generators of Sp(4) in the Traditional Notation}
This Appendix has two different purposes.  First, in order to
maintain continuity, we give the expressions for the generators for
canonical transformations used in the traditional literature including
our own papers where the phase-space coordinates are
$(x_{1}, x_{2}, p_{1}, p_{2})$.  Second, we point that they do not
generate canonical transformations in the new coordinate system we
introduced in Sec.~\ref{canno}.

For a dynamical system consisting of two pairs of canonical variables
$x_{1}, p_{1}$ and $x_{2}, p_{2}$, we can use the coordinate variables
defined as
\begin{equation}\label{coord2}
\left(\eta _{1}, \eta _{2}, \eta _{3}, \eta _{4} \right) = \left(x_{1},
x_{2}, p_{1}, p_{2} \right) .
\end{equation}
Then the transformation of the variables from
$\eta _{i}$ to $\xi _{i}$ is canonical if
\begin{equation}\label{symp2}
M J \tilde{M} = J ,
\end{equation}
where
$$
M_{ij} = {\partial \over \partial \eta _{j}}\xi _{i},
$$
\noindent and
$$
J = \pmatrix{0&0&1&0\cr0&0&0&1\cr-1&0&0&0\cr0&-1&0&0} .
$$

We used this coordinate system in our earlier papers, and the
transformation from the coordinate system of Eq.(\ref{coord1}) to
this coordinate system is straight-forward.  The generators of the
$SP(4)$ group are
$$
L_{1} = {i\over 2}\pmatrix{0&\sigma _{1} \cr -\sigma _{1} & 0}, \qquad
L_{2} = {1\over 2}\pmatrix{\sigma _{2} & 0 \cr 0 & \sigma _{2}} ,
$$

\begin{equation}\label{eq.4}
L_{3} = {i\over 2}\pmatrix{0 & \sigma _{3} \cr -\sigma _{3} & 0} , \qquad
S_{3} = {i\over 2}\pmatrix{0&-I\cr I&0} .
\end{equation}
The following six symmetric generators anticommute with J.
$$
K_{1} = {i\over 2}\pmatrix{0 & \sigma _{3} \cr \sigma _{3} & 0 }, \qquad
K_{2} = {i\over 2}\pmatrix{I&0\cr 0& -I} , \qquad
K_{3} = -{i\over 2}\pmatrix{0 & \sigma _{1} \cr \sigma _{1} & 0} ,
$$

\noindent and
\begin{equation}\label{eq.5}
Q_{1} = {i\over 2}\pmatrix{-\sigma _{3} & 0 \cr 0 & \sigma _{3}}, \qquad
Q_{2} = {i\over 2}\pmatrix{0&I \cr I&0} , \qquad
Q_{3} = {i\over 2}\pmatrix{\sigma _{1} & 0 \cr 0 & -\sigma _{1}} .
\end{equation}

These generators satisfy the set of commutation relations for $Sp(4)$
given in Sec.~\ref{canno} :
$$
[L_{i}, L_{j}] = i\epsilon _{ijk} L_{k} ,\qquad
[L_{i}, K_{j}] = i\epsilon _{ijk} K_{k} , \qquad
[K_{i}, K_{j}] = [Q_{i}, Q_{j}] = -i\epsilon _{ijk} L_{k} ,
$$
$$
[L_{i}, S_{3}] = 0 ,  \qquad  [K_{i}, Q_{j}] = -i\delta _{ij} S_{3} ,
$$
\begin{equation}
[L_{i}, Q_{j}] = i\epsilon _{ijk} Q_{k} , \qquad
[K_{i}, S_{3}] =  -iQ_{i} , \qquad [Q_{i}, S_{3}] = iK_{i} .
\end{equation}

These generators indeed satisfy the Lie algebra for the $Sp(4)$ group.
However, the above J matrix is different from the J matrix of
Eq.(\ref{JKQ}), and the above set of generators is different from the
set given in Eqs.~(\ref{rotKQ}) and (\ref{sqKQ}).  This means that the
above generators form a $Sp(4)$-like subgroup different from the one
discussed in Secs.~\ref{canno} and \ref{isomor}.

In the new coordinate system, the above $J$ matrix is proportional to
$L_{2}$.  The remaining rotation generators are $S_{1}, S_{2}, S_{3}$,
and the boost generators are $K_{3}, Q_{3}, G_{3}$ and
$K_{1}, Q_{1}, G_{1}$.  This $Sp(4)$ subgroup is like one of the three
$O(2,3)$-like subgroup of $O(3,3)$.  Because the operators $G_{3}$ and
$G_{1}$ are noncanonical, this is not a canonical subgroup.

\end{appendix}

\end{document}